\documentclass[aps,prl,twocolumn,superscriptaddress]{revtex4}

\usepackage{graphicx}
\usepackage{amsmath}

\begin{document}

\title{Resolution of the mystery of counter-intuitive photon correlations in \\
far off-resonance emission from a quantum dot-cavity system}

\author{Martin Winger}
\author{Thomas Volz}
\affiliation{Institute of Quantum Electronics, ETH Zurich, 8093 Zurich, Switzerland}
\author{Guillaume Tarel}
\author{Stefano Portolan}
\affiliation{Institute of Theoretical Physics, Ecole Polytechnique F\'ed\'erale de Lausanne EPFL, CH-1015 Lausanne, Switzerland}
\author{Antonio Badolato}
\affiliation{Department of Physics and Astronomy, University of
Rochester, Rochester, NY 14627, USA }
\author{Kevin J.\ Hennessy}
\author{Evelyn L.\ Hu}
\affiliation{California NanoSystems Institute, University of California, Santa Barbara, California 93106, USA}
\author{Alexios Beveratos}
\affiliation{CNRS - Laboratoire Photonique et Nanostructures, Route de Nozay, F-91460 Marcoussis, France}
\author{Jonathan Finley}
\affiliation{Walter Schottky Institut, Am Coulombwall 3, D-85748
Garching, Germany}
\author{Vincenzo Savona}
\affiliation{Institute of Theoretical Physics, Ecole Polytechnique F\'ed\'erale de Lausanne EPFL, CH-1015 Lausanne, Switzerland}
\author{Ata\c c Imamo\u glu}
\affiliation{Institute of Quantum Electronics, ETH Zurich, 8093
Zurich, Switzerland}

\date{\today}
\begin{abstract}
In a coupled quantum-dot nano-cavity system, the photoluminescence from an
off-resonance cavity mode exhibits strong quantum correlations with the quantum
dot transitions, even though its autocorrelation function is classical. Using new
pump-power dependent photon-correlation measurements, we demonstrate that this
seemingly contradictory observation that has so far defied an explanation stems
from cascaded cavity photon emission in transitions between excited multi-exciton
states. The mesoscopic nature of quantum dot confinement ensures the presence of a
quasi-continuum of excitonic transitions part of which overlaps with the cavity
resonance.

\end{abstract}


\maketitle

A quantum dot (QD) coupled to a photonic crystal cavity provides a promising
system for studying cavity quantum-electrodynamics (QED) in the solid state
\cite{Khitrova:2006,Gerard:1998}. In contrast to their atom-based counterparts,
these systems exhibit features that arise from their complex environment. A common
effect that surfaced in previous experiments is strong off-resonant emission of a
cavity mode (CM) containing one or multiple QDs. Photon correlation measurements
revealed that the cavity-mode emission is anti-correlated with the QD excitons at
the level of single quanta, proving that \emph{cavity feeding} is mediated solely
by a single QD \cite{Hennessy:2007,Kaniber:2008}. Surprisingly, however, the
photon stream emitted by the far off-resonant CM did not show any significant
quantum correlations. Previous experimental
\cite{Suffczynski:2009,Ates:2009,Ota:2009} and theoretical
\cite{Hughes:2009,Yamaguchi:2008} investigations have focused on explaining cavity
feeding in terms of dephasing of the QD excitons mediated either by coupling to
acoustic phonons or to free carriers. However, all of the attempts to describe
cavity feeding using Markovian dephasing of the fundamental exciton line fail to
explain the above mentioned photon correlation signatures that appear to be true
for all studied QD cavity-QED systems.

In this Letter, we unequivocally demonstrate that the far off-resonant excitation of the CM is solely due to the
mesoscopic nature of quantum dot confinement, which in turn leads to an energetically broad cascaded emission of the
QD. In this setting, cavity feeding and its photon correlation signatures can be regarded as an intrinsic feature of
QD-cavity systems that arises from the complicated QD multi-exciton level structure. We carry out pump-power dependent
photoluminescence (PL) as well as photon auto- and cross-correlation measurements on a nano-structure incorporating a
single QD embedded in a photonic crystal (PC) defect cavity \cite{Hennessy:2007}. To explain our experimental
observations, we develop a new theoretical model for the QD-cavity system, perform numerical calculations of its
semi-classical dynamics and compare its predictions with the new experimental findings. While a quantitative
comparison between numerical and experimental results is intrinsically difficult, the qualitative agreement we achieve
is excellent. In particular, the unusual correlation features found experimentally are naturally reproduced by the
model and the simulations.

Before proceeding, we remark that acoustic phonons should contribute to cavity feeding \cite{Hohenester2009} when the
CM-exciton detuning does not exceed a cutoff energy determined by the inverse QD size, typically in the range of 1 meV
(0.7~nm). In contrast, cavity feeding has experimentally been observed for detunings ranging from $+10~\mathrm{meV}$
($-7~\mathrm{nm}$) to $-45~\mathrm{meV}$ ($+32~\mathrm{nm}$) (i.e. $\sim70$--$300$ CM linewidths). Our focus in this
Letter lies on this far off-resonance cavity emission phenomenon, which cannot be explained using (Markovian)
pure-dephasing or phonon-assisted processes.

\begin{figure}
    \includegraphics[width = 86mm]{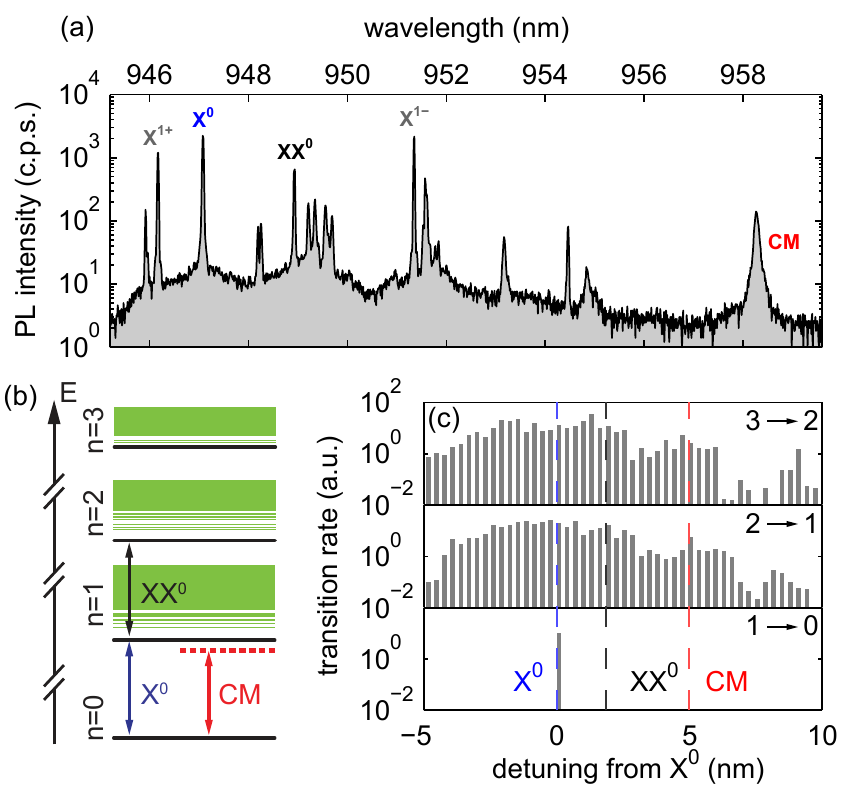}
    \caption{\label{Fig1} (color online) QD background emission and level structure.
        (a) Typical QD PL spectrum on a semi-logarithmic scale. Besides the well-known discrete QD lines, the weak background responsible for cavity feeding is visible.
        (b) Calculated energy-level diagram ($n\le 3$) of a neutral QD assuming a truncated parabolic in-plane confinement potential (energy separations between manifolds not to scale).
        (c) Calculated transition rates between states of different manifolds. States from $n \geq 2$ manifolds can decay into a large number of final states, thus forming a quasi-continuum of allowed transitions that can feed the CM.
        }
\end{figure}

A key aspect of cavity-feeding can be  identified from typical PL spectra of a single QD coupled to a PC cavity.
Figure~\ref{Fig1}(a) displays a PL spectrum of such a device on a semi-logarithmic scale with a cavity-mode frequency
tuned about $15.3~\text{meV}$ ($10.9~\mathrm{nm}$) red of the $X^0$ transition using nitrogen deposition
\cite{Srinivasan:2007,Mosor:2005}. The spectrum was obtained in a liquid-helium flow cryostat by pumping the device at
around $838~\text{nm}$, just below the GaAs bandgap. Even though the QD is driven well below saturation, the CM
emission is clearly visible together with the dominant QD PL lines that originate mainly from the neutral exciton
$X^0$, the negatively ($X^{1-}$) and positively ($X^{1+}$) charged trions, and the biexciton ($XX^0$) transition
\cite{Winger:2008}. We remark here that for our devices and excitation conditions, the magnitude of $X^0$, $X^{1-}$,
and $X^{1+}$ are comparable, suggesting that the QD is equally likely to be neutral or charged by a single excess
electron or a hole. In addition to these discrete lines, we observe a much broader single-QD background about two
orders of magnitude weaker in intensity. The part of this background that is resonant with the CM is Purcell enhanced
and leads to the off-resonant cavity luminescence. We emphasize that the principal aim of this Letter is to explain
the origin of this single-QD background and how it leads to the observed striking photon correlation signatures.

To understand this background  emission, we consider the full
excitation spectrum of electron and hole motional states in the QD.
While the narrow QD lines originate from states in which the
carriers occupy lowest energy single-particle states, there is a
large variety of higher excited states for which e.g.\ a carrier is
excited to a p- or d-shell state. This leads to an excitation
spectrum consisting of a series of QD manifolds separated
approximately by the band-gap energy.  We calculate the
multi-excitonic eigenstates up to four electron-hole pairs
\cite{Biolatti:2002,Barenco:1995} by diagonalizing the Coulomb
Hamiltonian, starting from a basis of 18 electron and 32 hole states
including spin. The result of the calculation is displayed in
Fig.~\ref{Fig1}(b). In addition, it has recently been demonstrated
that higher energy orbital states in an excitation manifold $n$
(with $n \ge 2$ electron-hole pairs) are subject to strong
hybridization with the wetting-layer continua \cite{Karrai:2004}.
This leads to the formation of a continuum of QD excited states for
each $n$.
\begin{figure}
        \includegraphics[width = 86mm]{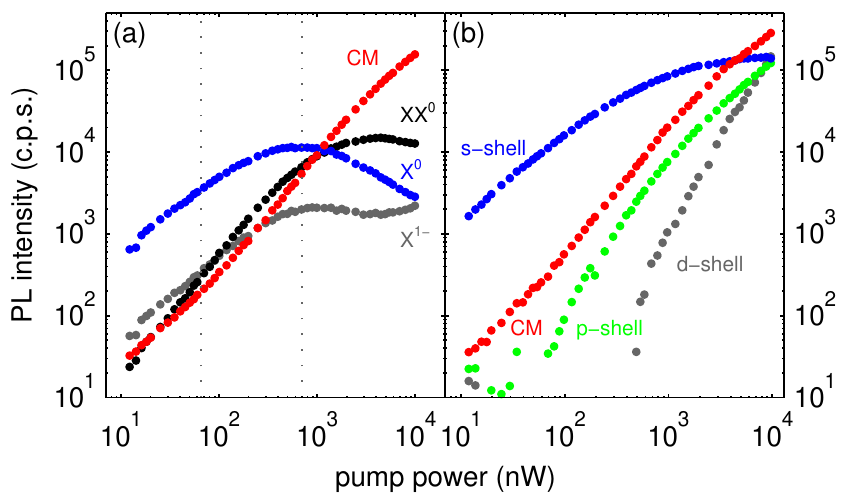}
        \caption{\label{Fig2} (color online) Off-resonant cavity emission as a function of pump power.
            (a) Pump-power dependence of CM, $X^0$, $X^{1-}$, and $XX^0$ emission. The data were obtained by integrating PL intensities for different excitation powers in $\approx\! 0.3~\text{nm}$ wide windows around the different QD lines indicated in Fig~1(a). The CM exhibits complex dynamics due to cavity feeding from higher-lying multi-excitonic states.
            (b) Comparison of CM emission with the emission from the different QD shells. The CM dynamics is clearly dominated by the QD s-shell for powers below $500~\text{nW}$, while at higher powers the p-shell takes over.
        }
\end{figure}
In this more accurate picture for the QD spectrum, the $X^0$ line
corresponds to a transition from the lowest energy $n=1$ manifold
state to the ground state ($1\rightarrow0$), whereas $XX^0$ decay
relates to a transition between the lowest energy states of the
$n=2$ and $n=1$ manifolds ($2\rightarrow1$) [Fig.~\ref{Fig1}(b)].
Given the large variety of initial and final states, $2\rightarrow1$
(and higher manifold) transitions merge to a quasi-continuum and
give rise to QD background emission that feeds the CM \footnote{We
remark here that stability of triplet states in the $n \geq 2$
manifolds against LO-phonon relaxation is likely to play a key role
in enhancing cavity feeding from the excited orbital states of the
corresponding manifold.}. Figure~\ref{Fig1}(c) shows calculated
transition rates between states of different manifolds. For initial
states in manifolds $n \geq 2$, decay can occur over a wide spectral
range, while decay from $n=1$ states leads to a series of discrete
emission lines, with the first excited transition
$\sim30~\mathrm{meV}$ ($21~\mathrm{nm}$) blue detuned from the $X^0$
line (not visible in Fig.~\ref{Fig1}(c)). We note that similar
excitation spectra exist for a charged QD. In contrast to the $X^0$
level spectrum, here also the $n=0^{\pm}$ manifolds consist of
multiple states owing to the excited confined motional states of the
extra charge. Hence, for charged QDs cavity feeding can also occur
for $1^\pm\rightarrow0^\pm$ transitions.

Due to the excited-state nature of cavity feeding for a neutral QD,
we expect a super-linear pump-power dependence of the CM emission.
In Fig.~\ref{Fig2}(a) we show the integrated intensities of the CM
together with those of the $X^0$, $X^{1-}$, and $XX^0$ lines as a
function of pump power in a log-log plot. As expected, the $X^0$ and
$X^{1-}$ lines follow an approximately linear power dependence below
saturation (which is defined through the emission maximum of the
respective line). A power-law fit to the $XX^0$ data gives an
exponent of $\sim\! 1.5$. In contrast, the dynamics of the CM is
more complex. For pump powers below $\sim\! 100~\mathrm{nW}$, the CM
follows a linear dependence, similar to the $X^0$ and the $X^{1-}$.
This observation is consistent with our model since for very low
pump powers we expect the charged $1^\pm\rightarrow0^\pm$ exciton
transitions to give the dominant contribution to cavity feeding.
When increasing the pump power, the behavior of the CM emission
becomes super-linear: in between the two vertical lines in
Fig.~\ref{Fig2}(a), the cavity displays the power dependence of the
$XX^0$ line. Furthermore, the cavity intensity increases far above
the saturation levels of both the $X^0$ and the $XX^0$. For powers
above $500~\text{nW}$, the cavity luminescence follows that of the
QD p-shell as clearly demonstrated in Fig.~\ref{Fig2}(b). Here we
recorded PL data over a wide spectral window covering the QD s-, p-
and d-shells and plot the total integrated intensity of the shell
emissions as a function of pump power. Since p- and d-shell emission
originate from higher-excitation manifolds of the QD
\cite{Bayer:2000}, these observations strongly support our model of
cavity feeding.

\begin{figure}
        \includegraphics[width = 86mm]{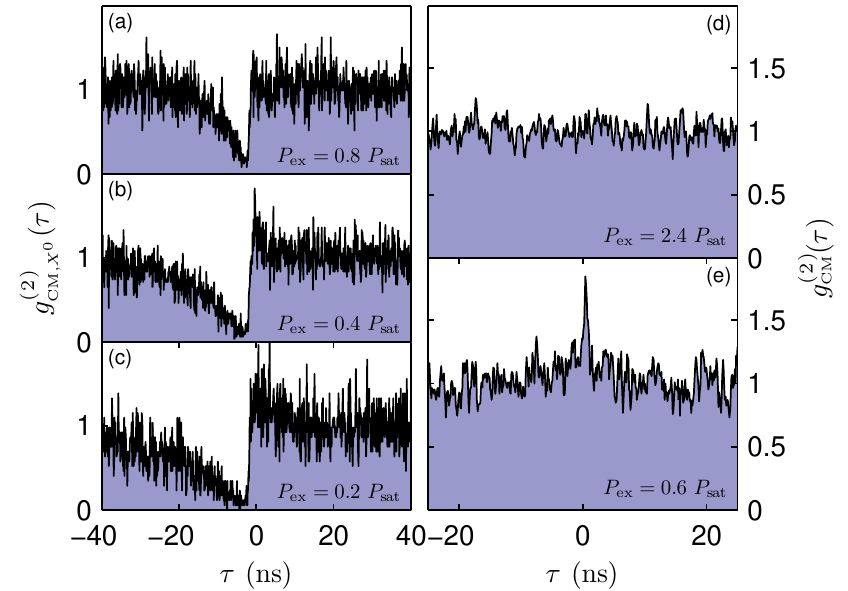}
        \caption{\label{Fig3} (color online) Experimental correlation histograms.
            (a)--(c) Cross-correlations between the CM and the $X^0$ for different pump powers. $\tau>0$ refers to $X^0$ emission upon detection of a cavity photon. The bunching feature for $\tau>0$ disappears when increasing the pump power, as expected from our model. The timescale for the anti-bunching feature for negative delays reflects the re-pump time of the system.
            (d)--(e) CM auto-correlations for two different pump powers. Below saturation, clear bunching at $\tau=0$ is visible which disappears above saturation, in agreement with the model.
        }
\end{figure}

As argued earlier, the most striking feature of cavity feeding  is
the peculiar interplay of quantum correlations observed in
off-resonant cavity PL \cite{Hennessy:2007,Kaniber:2008}. To obtain
a better understanding of these observations, we carried out
pump-power dependent photon correlation measurements that show new
features that were not visible in earlier measurements.
Figure~\ref{Fig3}(a)--(c) display normalized CM-$X^0$
cross-correlation curves for a mutual detuning of about
$10.3~\text{meV}$ ($7.3~\mathrm{nm}$) for three different pump
powers. The histograms were taken at pump powers of about (c)
$20\%$, (b) $40\%$ and (a) $80\%$ of the saturation power
$P_{sat}\approx 600~\text{nW}$. For all three pump powers, there is
a strong suppression of coincidences for zero time delay. In other
words, CM and $X^0$ emission do not occur simultaneously, which
follows naturally from our model since cavity feeding arises from QD
transitions other than the $X^0$. Moreover, we observe an abrupt
turn-on of $X^0$ luminescence upon detection of a cavity photon,
reflected by the bunching feature for $\tau>0$. This can easily be
explained by the fact that for pump powers below saturation a
significant portion of cavity photons is emitted in $2\rightarrow1$
transitions. In this case, detection of a cavity photon projects the
QD into the $n=1$ manifold, leading to a higher-than-average
probability for $X^0$ emission. When increasing the pump power, the
average QD population is higher, such that the detection of a cavity
photon does not increase the conditional probability that the QD is
in the $n=1$ manifold. This is manifested in a reduction and
eventually the disappearance of the bunching peak for higher powers,
as clearly visible in Figs.~\ref{Fig3}(a) and (b).

Surprisingly, in our previous study the cavity photons did not show
any significant quantum correlations \cite{Hennessy:2007} even
though the cavity is fed by a single quantum emitter. To gain
further insight, we measure the second-order auto-correlation
function $g_{\text{CM}}^{(2)}(\tau)$ of the cavity luminescence for
different pump powers. Here, the cavity is detuned by
$15.3~\text{meV}$ ($10.9~\mathrm{nm}$) to the red of the $X^0$.
Figure~\ref{Fig3}(d) and (e) display the outcome for two different
regimes of pump powers.
While for a pump power $P_{\text{ex}}\approx 0.6\ P_{\text{sat}}$
[Fig.~\ref{Fig3}(e)]  the cavity emission exhibits clear bunching
for $\tau = 0$, well above saturation the bunching disappears and
the correlations are purely Poissonian [Fig.~\ref{Fig3}(d)]. We
argue that the observation of bunching, not observed in earlier
single QD experiments, provides a strong indication of cavity
feeding from QD excited states as predicted by our model: when the
system is driven well below saturation, the likelihood of finding
the QD in a $n\geq2$ manifold is very small. Detection of a cavity
photon however, increases the conditional probability of finding the
QD in a higher manifold, implying in turn that a second
photon-emission event is more likely than on average. In contrast,
for higher pump powers the average population of excited states is
higher and a detection of a cavity photon does not increase the
likelihood of a second photon detection event, since the emission of
a cavity photon can arise from transitions between any two
neighboring excitation manifolds ($n\rightarrow n-1$ with
$n=2,3,\dots$). In this regime, cascaded cavity-photon emission
leads to Poissonian statistics.

\begin{figure}
        \includegraphics[width = 86mm]{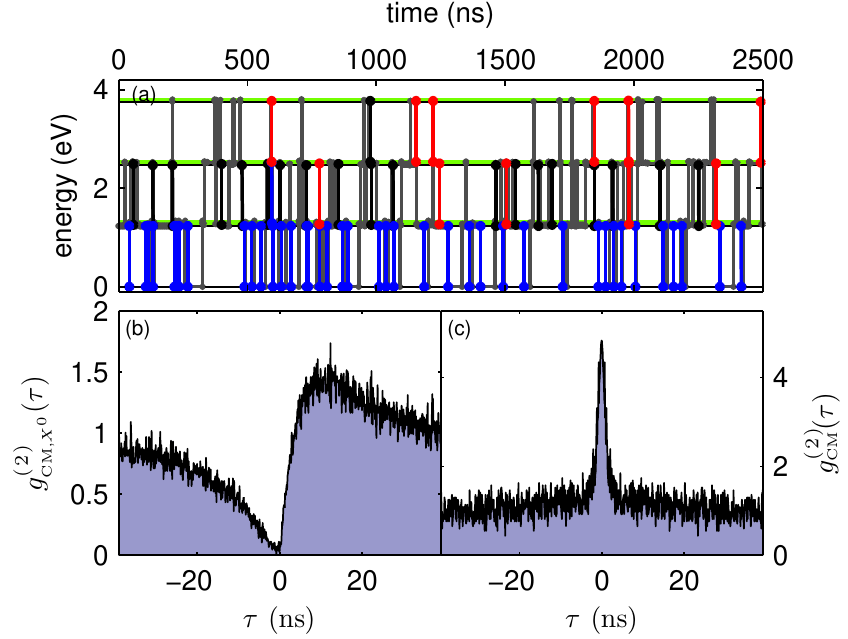}
        \caption{\label{Fig4} (color online) Simulation of cavity-feeding dynamics.
            (a) Monte-Carlo random-walk of excitation and photon-emission events (gray lines). Transitions resonant with the CM, $X^0$, and $XX^0$ lines are depicted in red, blue, and black, respectively.
            (b) Extracted CM-$X^0$ cross-correlation histogram reproducing the characteristic features observed in the experiment (compare Fig.~\ref{Fig3}(c)).
            (c) Extracted CM auto-correlation curve, exhibiting strong bunching at zero time delay in accordance with the experimental result of Fig.~\ref{Fig3}(e).
        }
\end{figure}

In order to demonstrate the consistency of our experimental findings
and our model, we determine the transition rates between the
calculated eigenstates by considering coupling to LO phonons, cavity
coupling and spin-flip processes. The dynamics of the system is then
studied by performing a Monte-Carlo random walk for a given
excitation rate \cite{Jacoboni:1983,Grundmann:1997}. A typical
result depicting the quantum jumps occurring in the system is shown
in Fig.~\ref{Fig4}(a). The events taking place at the cavity-mode
frequency are indicated by red lines, whereas $X^0$ and $XX^0$
events are shown in blue and black, respectively. Clearly, all
cavity photons here are emitted in $2\rightarrow1$ and
$3\rightarrow2$ transitions; in contrast, there are no cavity
photons emitted in $1\rightarrow0$ transitions, since there cavity
feeding is energetically not allowed, in perfect agreement with the
arguments given above. We remark that our simulation does not
include cavity feeding from charged excitonic compounds. However,
including charged exciton states would not affect the general
validity of the model or our findings.

>From the simulated Monte-Carlo random walk, we extract both
cross-correlation and auto-correlation histograms which are
displayed in Fig.~\ref{Fig4}(b) and (c).
In the simulation a CM detuning of $7~\mathrm{meV}$
($5~\mathrm{nm}$) red of the $X^0$ was assumed \footnote{The
truncated continuum of wetting-layer states leads to non-physical
results when going to larger detunings.}. In general, the
experimental observations of Fig.~\ref{Fig3} are well reproduced by
the simulations. The cross-correlation in Fig.~\ref{Fig4}(b)
exhibits the two main features already observed in the experiment:
anti-bunching at zero time delay and a bunching peak for positive
times \footnote{Here the bunching feature for $\tau>0$ is delayed
compared to the experimental trace, which could partly be due to the
simplified model used for the spin-flip process.}.
Figure~\ref{Fig4}(c) shows a cavity auto-correlation trace for a
pump power below saturation. As expected from our qualitative
arguments, the curve shows strong bunching with
$g^{(2)}_{\text{CM}}(0)\approx\!4.8$. In the experiment, however, we
only measure $g^{(2)}_{\text{CM}}(0)\approx\!1.8$, which we
attribute to two effects: first, the experimental pump power is
difficult to relate to the theoretical excitation rate and
$g^{(2)}_{\text{CM}}(\tau)$ strongly depends on the pump power.
Second, in the experiment cavity feeding can additionally arise from
charged $1^\pm\rightarrow0^\pm$ transitions. Consequently, we expect
a reduction of bunching, since here an $n\geq2^\pm$ manifold
excitation is not required for cavity feeding.

In conclusion, we have developed a model for cavity feeding based on
cascaded QD emission from excited states of multi-exciton manifolds.
In particular, the peculiar photon-correlation features observed
experimentally can fully be accounted for in our model. Moreover,
the  model is valid for a large range of detunings --- particularly
in the far-detuned case for which acoustic phonon mediated feeding
cannot play a role. The work presented here allows for a more
refined understanding of QD-based implementations of cavity QED and
the limitations of future spin-photon interfaces such devices are
predicted to realize. Moreover, it could have strong implications
for experiments on lasing \cite{Strauf:2006,Nomura:2009} in such
systems. An interesting extension of the present work could be the
investigation of feeding in charge-controlled devices
\cite{Hofbauer:2007}.

This work is supported by NCCR Quantum Photonics (NCCR QP), research
instrument of the Swiss National Science Foundation (SNSF).

\begin{thebibliography}{99}

\bibitem{Khitrova:2006}
G.\ Khitrova \textit{et al.}, Nature Physics \textbf{2}, 81 (2006).

\bibitem{Gerard:1998}
J.\ M.\ Gerard \textit{et al.}, Phys.\ Rev.\ Lett.\ \textbf{81},
1110 (1998).

\bibitem{Hennessy:2007}
K.\ Hennessy \textit{et al.}, Nature (London) \textbf{445}, 896
(2007).

\bibitem{Kaniber:2008}
M.\ Kaniber \textit{et al.}, Phys.\ Rev.\ B \textbf{77}, 161303(R)
(2008).

\bibitem{Suffczynski:2009}
J.\ Suffczynski \textit{et al.}, Phys.\ Rev.\ Lett.\ \textbf{103},
027401 (2009).

\bibitem{Ates:2009}
S.\ Ates \textit{et al.}, arXiv:quant-ph/0902.3455 (2009).

\bibitem{Ota:2009}
Y.\ Ota \textit{et al.}, arXiv:cond-mat/0908.0788 (2009).

\bibitem{Hughes:2009}
S.\ Hughes and P.\ Yao, Opt.\ Expr.\ \textbf{17}, 3322 (2009).

\bibitem{Yamaguchi:2008}
M.\ Yamaguchi and T.\ Asano and S.\ Noda, Opt.\ Expr.\ \textbf{16},
18067 (2008).

\bibitem{Hohenster2009}
U. Hohenester (private communication).

\bibitem{Srinivasan:2007}
K.\ Srinivasan and O.\ Painter, Appl.\ Phys.\ Lett.\ \textbf{90},
031114 (2007).

\bibitem{Mosor:2005}
S.\ Mosor \textit{et al.}, Appl.\ Phys.\ Lett.\ \textbf{87}, 141105
(2005).

\bibitem{Winger:2008}
M.\ Winger \textit{et al.}, Phys.\ Rev.\ Let.\ \textbf{101}, 226808
(2008).

\bibitem{Biolatti:2002}
E.\ Biolatti \textit{et al.}, Phys.\ Rev.\ B \textbf{65}, 075306
(2002).

\bibitem{Barenco:1995}
A.\ Barenco and M.\ A.\ Dupertuis, Phys.\ Rev.\ B \textbf{52}, 2766
(1995).

\bibitem{Karrai:2004}
K.\ Karrai \textit{et al.}, Nature \textbf{427}, 135 (2004).

\bibitem{Bayer:2000}
M.\ Bayer \textit{et al.}, Nature \textbf{405}, 923 (2000).

\bibitem{Jacoboni:1983}
C.\ Jacoboni and L.\ Reggiani, Rev.\ Mod.\ Phys.\ \textbf{55}, 645
(1983).

\bibitem{Grundmann:1997}
M.\ Grundmann and D.\ Bimberg, Phys.\ Rev.\ B \textbf{55}, 9740
(1997).

\bibitem{Strauf:2006}
S.\ Strauf \textit{et al.}, Phys.\ Rev.\ Lett.\ \textbf{96}, 127404
(2006).

\bibitem{Nomura:2009}
M.\ Nomura \textit{et al.}, arXiv:cond-mat/0906.4181 (2009).

\bibitem{Hofbauer:2007}
F.\ Hofbauer \textit{et al.}, Appl.\ Phys.\ Lett.\ \textbf{91},
201111 (2007).
\end{thebibliography}

\end{document}